\documentclass[pre]{revtex4-2}
\usepackage{url}
\usepackage{dcolumn}
\usepackage{amssymb, mathtools, siunitx}
\usepackage[dvipdfmx]{graphicx}

\DeclareMathOperator{\erf}{erf} 

\begin{document}
\title{Analysis of size distributions of fictional characters: from \textit{Pok\'emon} to \textit{Godzilla}}

\author{Ken Yamamoto}
\affiliation{Department of Physics and Earth Sciences, Faculty of Science, University of Ryukyus, Nishihara, Okinawa 903--0213, Japan}

\begin{abstract}
In this letter, the size (height and weight) of fictional characters in animations, superhero series, movies, and other media is studied.
We find that the distributions of character height and weight approximately follow lognormal distributions in common to five selected works.
We propose a mechanism governing this lognormal behavior based on the principle of maximum entropy and the Weber-Fechner law.
Moreover, we provide a comparison to the size distributions of real animals.
Although the size distributions of fictional characters and real animals are both lognormal, the distributions are essentially different, particularly in the scaling between height and weight.
\end{abstract}

\maketitle

Size distribution is an important tool in analyzing complex systems.
We can compute size distributions simply by collecting and organizing data.
In general, obtaining precise microscopic dynamics in complex systems is difficult, and an effective approach for it is a phenomenological description~\cite{Castellano, Buchanan}.
The properties of size distributions is useful in constructing phenomenological models.
In many cases, size distributions follow heavy-tailed probability distributions whose decay in the tail is slower than the exponential functions; thus, the mean, variance, or higher moments diverge~\cite{Foss}.
The emergence of heavy-tailed distributions indicates the existence of elements much larger than the ones of typical size.
Further, numerous theoretical mechanisms have been proposed to obtain heavy-tailed distributions, such as power-law and lognormal distributions~\cite{Mitzenmacher, Newman}.

Generally, the size of the biological organisms typically follows lognormal distribution~\cite{Koch, Limpert, Yamamoto, Koyama}.
This is a heavy-tailed distribution where the logarithm of the size is distributed normally~\cite{Kobayashi}.
Theoretically, the lognormal behavior of organisms is simply explained by the multiplicative stochastic process
\begin{equation}
X_{t+1} = R_t X_t,
\label{eq1}
\end{equation}
where $R_t$ is a random growth rate.
In the simplest case, when $R_1, R_2,\ldots$ are independently and identically distributed, $\ln X_t$ approximately follows a normal distribution for sufficiently large $t$ (because of the central limit theorem), and $X_t$ follows a lognormal distribution.
This model for the lognormal distribution dates back to Kolmogorov~\cite{Kolmogorov}.

In this study, we analyzed the sizes (height and weight) of \textit{fictional} creatures, instead of real organisms.
They are characters in creations such as animations, movies, and comics whose sizes are arbitrarily determined by their designers.
Nevertheless, we find that size distributions for different works exhibit a common lognormal behavior.
(A preliminary result was already reported~\cite{Yamamoto2017}.)
We propose a simple mechanism that produces this lognormal distribution based on the principle of maximum entropy and the Weber-Fechner law regarding the perception of numbers.
Moreover, we investigate the similarities and differences between fictional characters and real animals from the perspective of size distribution.

To analyze the size distribution, it is desirable to select works in which the sizes of a large number of characters are officially determined.
We analyze the following five groups of characters:
\begin{enumerate}
\item Pok\'emon (creatures) from the \textit{Pok\'emon} series (video games, animations, and other media)
\item Zords (robots and vehicles) from the \textit{Power Rangers} series (American superhero television series)
\item Giant monsters from the \textit{Godzilla} series (Japanese movie series),
\item Villains (giant monsters and alien invaders) from the \textit{Ultraman} series (Japanese superhero television series)
\item Characters from the \textit{Dragon Ball} series (Japanese comics and animations of martial arts experts and alien warriors).
\end{enumerate}
We obtained the sizes of Pok\'emon creatures from the online official database \textit{Pok\'edex}~\cite{Pokemon} (up to Generation VII),
the sizes of Zords from the online database \textit{RangerWiki}~\cite{PowerRangers} (up to the 25th season, entitled \textit{Super Ninja Steel}),
and the sizes of characters in \textit{Dragon Ball} from the online database \textit{Dragon Ball Wiki}~\cite{Dragonball} in March 2019.
We referred to Japanese books~\cite{Godzilla, Ultraman} to obtain the character sizes in \textit{Godzilla} (including 29 movie titles from 1954 to 2016) and \textit{Ultraman} (\textit{Showa Ultraman} series aired in Japan from 1966 to 1981).
We only analyzed characters whose height and weight were both provided.

\begin{figure*}[t!]\centering
\raisebox{3.5cm}{a}
\includegraphics[scale=0.8]{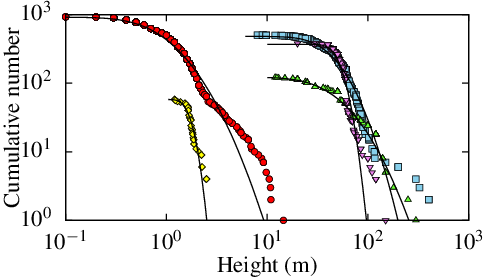}
\hspace{5mm}
\raisebox{3.5cm}{b}
\includegraphics[scale=0.8]{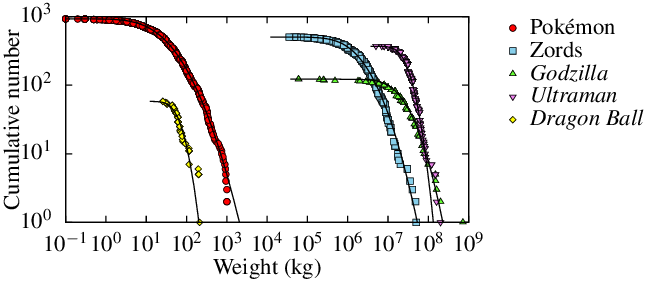}
\caption{
(Color online)
Cumulative height (a) and weight (b) distributions of Pok\'emon (circles), Zords in \textit{Power Rangers} (squares), \textit{Godzilla} monsters (upper triangles), \textit{Ultraman} villains (lower triangles), and \textit{Dragon Ball} characters (diamonds).
The curves present a lognormal fit for each group of characters.
}
\label{fig1}
\end{figure*}

Figure~\ref{fig1} shows the height (a) and weight (b) distributions in cumulative form.
Each curve is a lognormal distribution:
\[
F(x)=\frac{N}{2}\left(1-\erf\left(\frac{\ln x-\mu}{\sqrt{2}\sigma}\right)\right),
\]
where $N$ is the number of characters, $\erf$ is the Gauss error function, and $\mu$ and $\sigma$ are parameters.
The values of $N$, $\mu$ and $\sigma$ are listed in Table~\ref{tbl1}.
The height and weight distributions are both consistent with a lognormal distribution.
As shown in Fig.~\ref{fig1} a, approximately ten of the tallest characters in \textit{Pok\'emon} (circles), \textit{Power Rangers} (squares), and \textit{Ultraman} (lower triangles) deviate from the lognormal distribution.
In Fig.~\ref{fig1} b, only the heaviest character in \textit{Godzilla} (the rightmost upper triangle), SpaceGodzilla Flying Form with a weight of \SI{720000}{tons}~\cite{Godzilla}, deviates greatly from the lognormal curve.

\begin{table}[tb!]\centering
\caption{The number $N$ of characters and the lognormal parameters $\mu$ and $\sigma$ for the height and weight distributions in each work.}
\setlength{\tabcolsep}{2mm}
\begin{tabular}{lD{.}{}{0}D{.}{.}{2}D{.}{.}{2} D{.}{.}{2}D{.}{.}{2}}
\hline\hline
& & \multicolumn{2}{l}{Height} & \multicolumn{2}{l}{Weight}\\
\cline{3-4} \cline{5-6}
Title & \multicolumn{1}{l}{$N$} & \multicolumn{1}{l}{$\mu_\mathrm{H}$} & \multicolumn{1}{l}{$\sigma_\mathrm{H}$} & \multicolumn{1}{l}{$\mu_\mathrm{W}$} & \multicolumn{1}{l}{$\sigma_\mathrm{W}$}\\
\hline
\textit{Pok\'emon} & 927. & -0.05 & 0.74 & 3.34 & 1.40\\
\textit{Power Rangers} & 502. & 3.80 & 0.52 & 14.48 & 1.15\\
\textit{Godzilla} & 122. & 3.89 & 0.69 & 16.97 & 0.95\\
\textit{Ultraman} & 369. & 3.97 & 0.22 & 17.20 & 0.54\\
\textit{Dragon Ball} & 58. & 0.55 & 0.17 & 4.21 & 0.52\\
\hline\hline
\end{tabular}
\label{tbl1}
\end{table}

To investigate the lognormal distribution of character size in more detail, we applied quantile-quantile (Q-Q) plots~\cite{Thode}.
The $q$-quantile $Q_q$ (for $0\le q\le 1$) of a random variable $X$ is defined by $P(X\le Q_q)\ge q$ and $P(X\ge Q_q)\ge 1-q$.
The Q-Q plot is a scatter plot of the data and the corresponding theoretical quantiles of a certain reference probability distribution (usually a normal distribution).
If the points in the Q-Q plot are linearly aligned, the data are determined to follow the reference distribution well.

\begin{figure*}[tb!]\centering
\raisebox{2.9cm}{a1}\hspace{-4mm}
\includegraphics[scale=0.65]{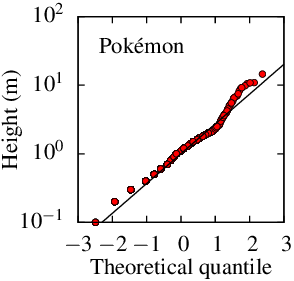}
\raisebox{2.9cm}{a2}\hspace{-2.5mm}
\includegraphics[scale=0.65]{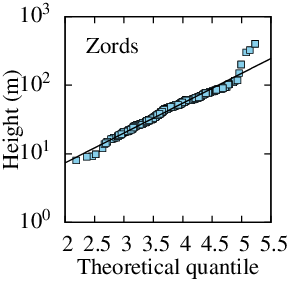}
\raisebox{2.9cm}{a3}\hspace{-2.5mm}
\includegraphics[scale=0.65]{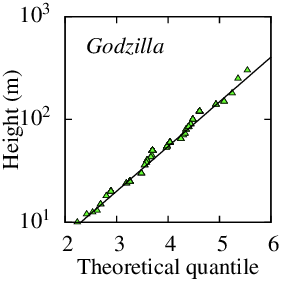}
\raisebox{2.9cm}{a4}\hspace{-2.5mm}
\includegraphics[scale=0.65]{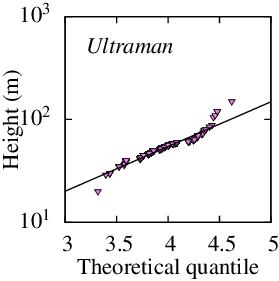}
\raisebox{2.9cm}{a5}\hspace{-2.5mm}
\includegraphics[scale=0.65]{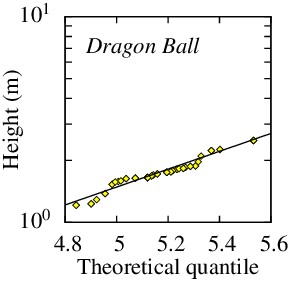}\\
\raisebox{2.9cm}{b1}\hspace{-4mm}
\includegraphics[scale=0.65]{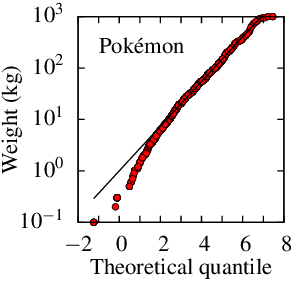}
\raisebox{2.9cm}{b2}\hspace{-3mm}
\includegraphics[scale=0.65]{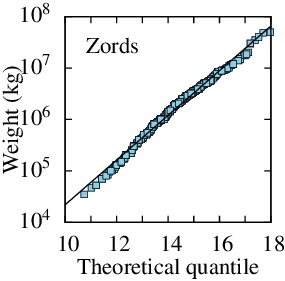}
\raisebox{2.9cm}{b3}\hspace{-3mm}
\includegraphics[scale=0.65]{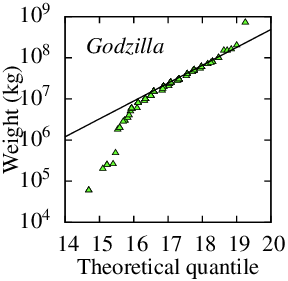}
\raisebox{2.9cm}{b4}\hspace{-3mm}
\includegraphics[scale=0.65]{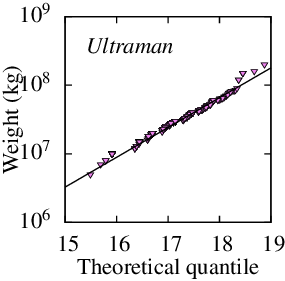}
\raisebox{2.9cm}{b5}\hspace{-3mm}
\includegraphics[scale=0.65]{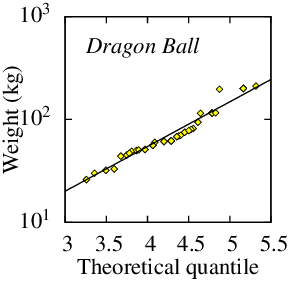}
\caption{
(Color online)
Q-Q plots of the height (a) and weight (b) of Pok\'emon (1), Zords (2), \textit{Godzilla} monsters (3), \textit{Ultraman} villains (4), and \textit{Dragon Ball} characters (5).
The diagonal line in each panel indicates the expected results for the case in which the sizes are lognormally distributed.
}
\label{figQQ}
\end{figure*}

Figure~\ref{figQQ} shows Q-Q plots for the height and weight of the characters, where log-transformed sizes are compared with the quantiles of a normal distribution.
(Note that a normal distribution for the log-transformed size indicates that the size is distributed lognormally.)
Among all of the characters, the weight of Zords (b2) and \textit{Ultraman} villains (b4) and the height of \textit{Godzilla} monsters (a3) agree very well with a lognormal distribution.
Some of the other Q-Q plots show that a few characters deviate from the lognormal distribution in large- or small-sized regions.
These deviations have a common tendency; large (tall or heavy) characters tend to have sizes greater than those expected by the lognormal distribution consistently (points in Fig.~\ref{figQQ} are above the solid diagonal lines), and small characters tend to be less than the lognormal distribution (points are below the diagonal line).
We surmise that the size of large characters can be overstated and those of small characters can be understated; for example, heavy-looking characters will be given excessively heavy weights to highlight their heaviness.
However, as a whole, the character size follows a lognormal distribution, particularly in the intermediate region.

Here, we propose a theoretical mechanism governing the common lognormal behavior of character sizes.
We introduce a random variable $X$ for the character size.
Because the characters are fictional, their sizes are free from physical, biological, or mechanical constraints.
Instead, the character sizes most likely reflect how humans interpret and perceive numbers.

The Weber-Fechner law states that the subjective intensity of a sensation (e.g., vision, hearing, or touch) is proportional not to the stimulus intensity but to its logarithm~\cite{Maor}.
In other words, this law implies that our mental number line has a logarithmic scale.
Notably, the Weber-Fechner law also governs our recognition of numbers~\cite{DehaeneBook}.
Furthermore, the Weber-Fechner law has been observed in neural~\cite{Dehaene} and mathematical~\cite{Portugal} studies.

A well-known example of the Weber-Fechner law is the perception of musical pitches.
We perceive that the pitch interval between the musical tones by two adjacent keys in piano is constant.
However, an octave rise in pitch corresponds to a doubling of frequency~\cite{Agmon}.
The frequency $f$ of the pitch $n$ octave above the frequency $f_0$ is written as $f=2^n f_0$.
Its logarithm, $\ln f=n\ln2+\ln f_0$, indicates that $\ln f$ increases at a constant rate $\ln 2$ with the increase in $n$.
Therefore, the logarithm of frequency matches the perception of pitches, which is a manifestation of the Weber-Fechner law.

A typical character size appears to exist for each work; for instance, the sizes of Pok\'emon are similar to those of human children.
Hence, in the simplest view, the size of a character is given by this typical size with fluctuations due to the individual differences.
Combined with the Weber-Fechner law, in which character sizes are mentally measured as $\ln X$, this condition is expressed as
\begin{equation}
E[\ln X]=\mu,\quad V[\ln X]=\sigma^2,
\label{eq2}
\end{equation}
where $E$ and $V$ represent the average and variance, respectively, and $\mu$ and $\sigma$ are associated with the typical character size and fluctuation, respectively.

According to the principle of maximum entropy~\cite{Jaynes}, when we have no knowledge about $X$ except for Eq.~\eqref{eq2},
the best probability density function $f(x)$ of $X$ is given by maximizing the entropy $-\int_0^\infty f(x)\ln f(x) dx$ under constraint~\eqref{eq2}.
We employ the method of Lagrange multipliers and set
\begin{align*}
H[f]&=-\int_0^\infty f(x)\ln f(x) dx - \lambda_0\left(\int_0^\infty f(x) dx-1\right)\\
&\quad-\lambda_1\left(\int_0^\infty f(x)\ln x dx -\mu\right)\\
&\quad-\lambda_2\left(\int_0^\infty f(x)(\ln x-\mu)^2 dx -\sigma^2\right).
\end{align*}
Multipliers $\lambda_1$ and $\lambda_2$ correspond to constraints shown in Eq.~\eqref{eq2}, and $\lambda_0$ is needed for the normalization $\int_0^\infty f(x) dx=1$.
To find $f$ such that the functional $H[f]$ takes a stationary value, we calculate
\[
\frac{\delta H}{\delta f}=-\ln f -1-\lambda_0-\lambda_1\ln x-\lambda_2(\ln x-\mu)^2=0,
\]
which directly leads to
\[
f(x)=e^{-1-\lambda_0}\frac{1}{x}\exp\left(-\lambda_2(\ln x-\mu)^2+(1-\lambda_1)\ln x\right).
\]
The exponent of the exponential function is a quadratic of $\ln x$; thus, $f$ is a lognormal-type function.
By determining the multipliers $\lambda_0, \lambda_1$, and $\lambda_2$, the function $f$ is found to be the probability density of the lognormal distribution with parameters $\mu$ and $\sigma$:
\[
f(x) = \frac{1}{\sqrt{2\pi\sigma^2}x}\exp\left(-\frac{(\ln x-\mu)^2}{2\sigma^2}\right).
\]
Therefore, the principle of maximum entropy combined with the Weber-Fechner law can yield a lognormal distribution.
Other mechanisms for lognormal distributions based on the Weber-Fechner law have been proposed for the length of user posts on the Internet~\cite{Sobkowicz},
the length of English words and the number of strokes in Chinese characters~\cite{Huang}, and the size of human tumors~\cite{Spratt}.

We find that the lognormal distribution is a standard distribution for fictional character sizes.
We also note that the above conclusion of lognormal distribution is derived from the simplest assumption.
Because character sizes are set arbitrarily, a designer can intentionally determine the character sizes that do not follow a lognormal distribution.
We believe that the lognormal distribution is natural, in the sense that the character-size distribution tends to be lognormal when the size is set without substantial thought.

\begin{figure}[tb!]\centering
\raisebox{2.6cm}{a}
\includegraphics[scale=0.6]{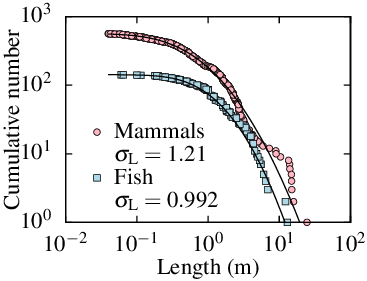}
\raisebox{2.6cm}{b}
\includegraphics[scale=0.6]{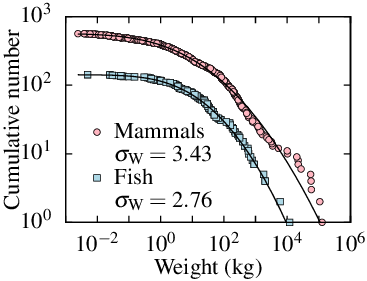}
\raisebox{2.6cm}{c}
\includegraphics[scale=0.6]{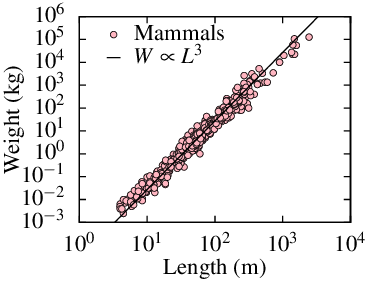}
\raisebox{2.6cm}{d}
\includegraphics[scale=0.6]{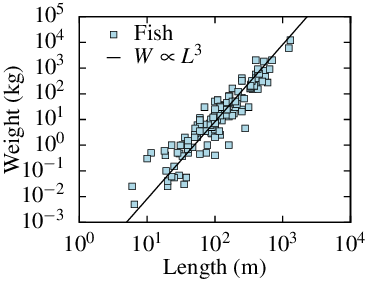}
\caption{
(Color online)
(a) Cumulative length distributions of 564 mammals (circles) and 124 fish (squares).
The solid curves are lognormal distributions, with $\mu_\mathrm{L}=-0.565$ and $\sigma_\mathrm{L}=1.21$ for mammals and $\mu_\mathrm{L}=0.051$ and $\sigma_\mathrm{L}=0.992$ for fish.
(b) Cumulative weight distributions of mammals and fish.
The lognormal parameters are $\mu_\mathrm{W}=1.61$ and $\sigma_\mathrm{W}=3.43$ for mammals and $\mu_\mathrm{W}=2.36$ and $\sigma_\mathrm{W}=2.76$ for fish.
(c) Correlation between the length and weight of mammals.
The solid line represents the scaling $W\propto L^3$ based on the principle of similitude.
(d) The length and weight of fish.
}
\label{fig3}
\end{figure}

We compare the size distributions of fictional characters and real animals.
Figures~\ref{fig3}~a and b show the length and weight distributions of 564 species of mammals (circles) and 124 species of fish (squares), obtained from a visual guide of animals~\cite{Burnie};
we included only species for which both the length and weight were given.
Although the circle points in Fig.~\ref{fig3} a and b deviate slightly from the solid lognormal fitting curves in the large-sized regions (the author does not know why the deviation occurs), the overall distributions can be approximated by a lognormal distribution.

Equation~\eqref{eq1} indicates that the lognormality of real animals of a single species arises from the randomness in their growth.
However, this equation cannot be directly applied to the sizes of different species, as shown in Fig.~\ref{fig3} a and b.
Similarly, the lognormal life-span distribution of invertebrates has been reported~\cite{Kobayashi2006}.
The possibility of the existence of a multiplicative stochastic effect in evolutionary processes is indicated as the cause of this lognormality; however, this hypothesis has not been completely verified.
Although the principle behind the lognormal distribution in Fig.~\ref{fig3} a and b has not been accurately determined, it is evident that growth and evolution are not associated with the lognormality of fictional characters.
In fact, characters usually do not grow or age; child characters occasionally grow, but they do not grow continuously as described by Eq.~\eqref{eq1}.
Incidentally, the ``evolution'' of some Pok\'emon does not cause a multiplicative effect, because Pok\'emon can evolve, at most, only twice, which is insufficient to yield a lognormal distribution.
Therefore, although both fictional characters and real animals exhibit lognormal size distribution, their principles of generation are completely different.

One might assume that the lognormal behavior of fictional characters arises from the imitation of real creatures, whose sizes are distributed lognormally; for instance, Pikachu in \textit{Pok\'emon} is modeled from a mouse~\cite{Pokemon}.
To determine whether this hypothesis is valid, we focus on the value of the lognormal fitting parameter $\sigma$.
Whereas $(\sigma_\mathrm{L}, \sigma_\mathrm{W})=(1.21, 3.43)$ for mammals and $(0.992, 2.76)$ for fish, the values of $(\sigma_\mathrm{H}, \sigma_\mathrm{W})$ for fictional characters are much smaller (see Table~\ref{tbl1}).
(Note that $\sigma_\mathrm{H}$ for character height is the counterpart of $\sigma_\mathrm{L}$ for animal length because height and length have linear dimensions.)
By comparing the parameter $\sigma$, the lognormality of characters is found to be independent of the size of real creatures.

The difference between the size of fictional characters and that of real animals is further clarified by the correlation between their weight and height (or length).
If animals had a homothetic body shape and a constant density, their weight $W$ and length $L$ would satisfy $W\propto L^3$, known as the principle of similitude~\cite{Thompson}.
As shown in Fig.~\ref{fig3} c and d, the scatter plots of the length and weight for mammals and fish are both consistent with the principle of similitude (solid lines).
Taking the logarithm, we have $\ln W=3\ln L+\text{constant}$.
Because the parameters $\sigma_{L}$ and $\sigma_{W}$ of the lognormal distribution represent the standard deviations of $\ln L$ and $\ln W$, respectively,
we have $\sigma_\mathrm{W}=3\sigma_\mathrm{L}$.
In other words, the ratio $\sigma_\mathrm{W}/\sigma_\mathrm{L}$ is a good estimate for the dimensionality of animals.
Indeed, for the data in Fig.~\ref{fig3}, $\sigma_\mathrm{W}/\sigma_\mathrm{L}=2.86$ for mammals and $\sigma_\mathrm{W}/\sigma_\mathrm{L}=2.78$ for fish, which are both close to three.

\begin{figure*}[tb!]\centering
\raisebox{2.9cm}{a}\hspace{-2.8mm}
\includegraphics[scale=0.65]{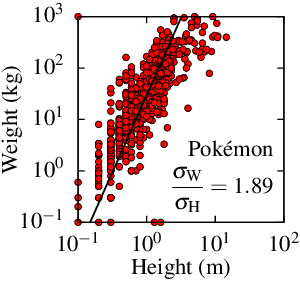}
\raisebox{2.9cm}{b}\hspace{-2mm}
\includegraphics[scale=0.65]{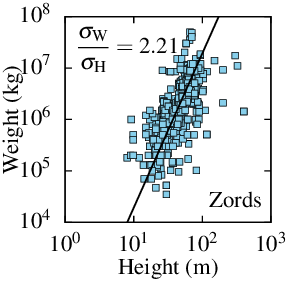}
\raisebox{2.9cm}{c}\hspace{-2mm}
\includegraphics[scale=0.65]{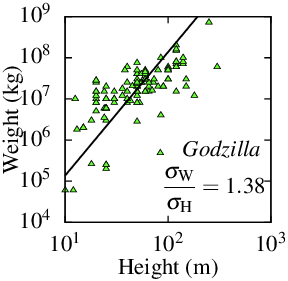}
\raisebox{2.9cm}{d}\hspace{-2mm}
\includegraphics[scale=0.65]{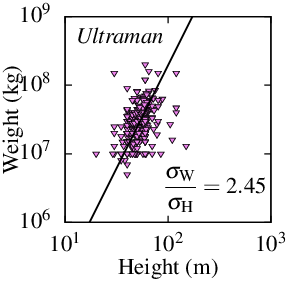}
\raisebox{2.9cm}{e}\hspace{-2mm}
\includegraphics[scale=0.65]{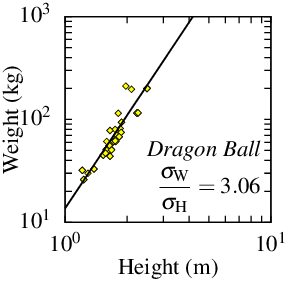}
\caption{
(Color online)
Scatter plots of the height and weight of Pok\'emon~(a), Zords~(b), \textit{Godzilla} monsters~(c), \textit{Ultraman} villains~(d), and \textit{Dragon Ball} characters~(e).
The solid line in each panel is based on the principle of similitude, $W\propto H^3$.
}
\label{fig4}
\end{figure*}

Scatter plots for the height $H$ and weight $W$ of characters are shown in Fig.~\ref{fig4}.
Only the \textit{Dragon Ball} characters (e) follow the scaling $W\propto H^3$ ($\sigma_\mathrm{W}/\sigma_\mathrm{H}=3.06$).
We believe that \textit{Dragon Ball} characters can easily be assigned realistic height and weight, because many of the characters are humans or human-shaped aliens.
The other plots, a--d, exhibit a roughly positive correlation, but the fluctuations are large compared with those of mammals and fish, as shown in Fig.~\ref{fig3}~c and d.
Thus, small but heavy characters and large but light characters are included.
The most extreme character is Cosmoem in \textit{Pok\'emon}~\cite{Pokemon}, which is \SI{0.1}{\metre} tall but \SI{999.9}{\kilogram} heavy.
Such unrealistic combinations of height and weight may help emphasize the peculiarity of a character.
The estimated $\sigma_\mathrm{W}/\sigma_\mathrm{H}$ of dimensionality is considerably smaller than three (close to two), in contrast to the finding for real animals.

In summary, the height and weight of fictional characters in animations, movies, and other media approximately follow a lognormal distribution,
even though character sizes are set arbitrarily for each work and are not associated with the size of real animals.
In this letter, we propose that this lognormal behavior can be explained by entropy maximization under the constraints where mean and variance of the size, determined by applying the Weber-Fechner law, are fixed.
This study implies that statistical regularity and commonality arise from the process of perception and generation of numbers by humans.
We expect this work to motivate further studies on the complex thought process and behavior of humans.

The present study was supported by the Hayao Nakayama Foundation for Science \& Technology and Culture (H29-B-41)
and by a Grant-in-Aid for Scientific Research (C) (19K03656) from the Japan Society for the Promotion of Science.

\end{document}